\documentclass[conference]{IEEEtran}
\IEEEoverridecommandlockouts
\usepackage{cite}
\usepackage{amsmath,amssymb,amsfonts}
\usepackage{algorithmic}
\usepackage{graphicx}
\usepackage{textcomp}
\usepackage{blindtext}
\usepackage{xcolor}
\def\BibTeX{{\rm B\kern-.05em{\sc i\kern-.025em b}\kern-.08em
    T\kern-.1667em\lower.7ex\hbox{E}\kern-.125emX}}
\makeatletter
\def\ps@IEEEtitlepagestyle{
	\def\@oddfoot{\mycopyrightnotice}
	\def\@evenfoot{}
}
\def\mycopyrightnotice{
	{\footnotesize
		\begin{minipage}{\textwidth}
			\centering
			Copyright~\copyright~ VTC Fall workshop'19. Personal use of this material is permitted. Permission from IEEE/VDE must be obtained for all other uses, in any current or future media, including reprinting/republishing this material for advertising or promotional purposes, creating new collective works, for resale or redistribution to servers or lists, or reuse of any copyrighted component of this work in other works.
		\end{minipage}
	}
}

\begin{document}

\title{Methodologies of Link-Level Simulator and System-Level Simulator for C-V2X Communication}
\IEEEpeerreviewmaketitle
\author{\IEEEauthorblockN{Donglin Wang, Raja R.Sattiraju, Anjie Qiu, Sanket Partani and Hans D. Schotten}
	\IEEEauthorblockA{\textit{University of Kaiserslautern} \\
		Kaiserslautern, Germany \\
		$\{$dwang,sattiraju,qiu,partani,schotten$\}$@eit.uni-kl.de}
}

\maketitle

\begin{abstract}
At the time of the development, standardization, and further improvement are vital to the modern cellular systems such as the next generation wireless communication (5G). Simulations are essential to test and optimize algorithms and procedures prior to their implementation process of the equipment manufactures. In order to evaluate system performance at different levels, accurate simulations of simple setups, as well as simulations of more complex systems via abstracted models are necessary. In this work, two new simulators for the sidelink Cooperative-Vehicle-to-Everything (C-V2X) communication have been implemented and carried out on both the physical layer (Link-Level (LL)) and network layer (System-Level (SL)). Detailed methodologies of the LL and SL simulators for C-V2X communication have been illustrated. In the LL simulator, we get the mapping curves of BLER and Signal-to-Noise-Ratio (SNR), which are used as a baseline for measuring the performance of the LL simulation. In addition, these mapping curves are used as the important Link-to-System (L2S) interfaces. The SL simulator is utilized for measuring the performance of cell networking and simulating large networks comprising of multiple eNBs and UEs. Finally, the simulation results of both simulators for C-V2X communication are presented, which shows that different objectives can be met by using LL or SL simulations types.
\end{abstract}


\section{Introduction}

      In wireless mobile communication, C-V2X communication can be used for information exchange among traffic participants. Now governments and organizations all over the world are trying to develop 5G to support C-V2X communication. However, the realistic measurement approaches for C-V2X communications are becoming costlier and more laborious in the real world. Thus, it’s extremely essential and inevitable to simulate the communication network to figure out the mutual information interactions of all involved elements. Especially, simulations are more efficient for gaining insights into the performance of both small-and-large scale scenarios. Therefore, simulations are carried out on both the physical layer (LL) and the network (SL) context [1]. If no computation limitations were to exist, both simulation types would be achievable by a single simulation tool. Realistically, each of the two simulation types has different objectives, which are met by them independently. The simulators are developed along with the standard-compliant Evolved Universal Terrestrial Radio Access (E-UTRA) / Long Term Evolution (LTE) 3rd Generation Partnership (3GPP) release-14 “Sidelink” radio technology and its integration into 5G network infrastructures [2]. Also, sidelink is known as PC5, which can facilitate the direct communication between the two ends of a C-V2X communication link. Nevertheless, a lot of simulations have been implemented (e.g. Vienna Simulators) which focus on the LTE downlink and uplink LL and SL simulators [3][4]. Following the ongoing discussion on 5G, we introduce the new 5G sidelink simulators. The simulations for our project are split into a LL simulator and a SL simulator respectively. At the beginning of this work,  we described the system model to facilitate sidelink C-V2X communication. Following that, we gave an overview of the capabilities and the general purposes of our LL simulator and SL simulator. Next, we introduced both architectures and key features of the LL and SL simulators in Sect. III and Sect. IV. Details for implementation of both simulations are described in Sect. V and Sect. VI. In Sect. VII, we got these simulation results for simulators of sidelink LTE C-V2X communications are obtained and the performances of this cellular wireless communications are evaluated. Moreover, comparisons have done in order to analyze the LL and SL simulators for different purposes requirements.

\section{System description}
	\begin{figure*}[htbp]
		\centering
		\includegraphics[width=\linewidth]{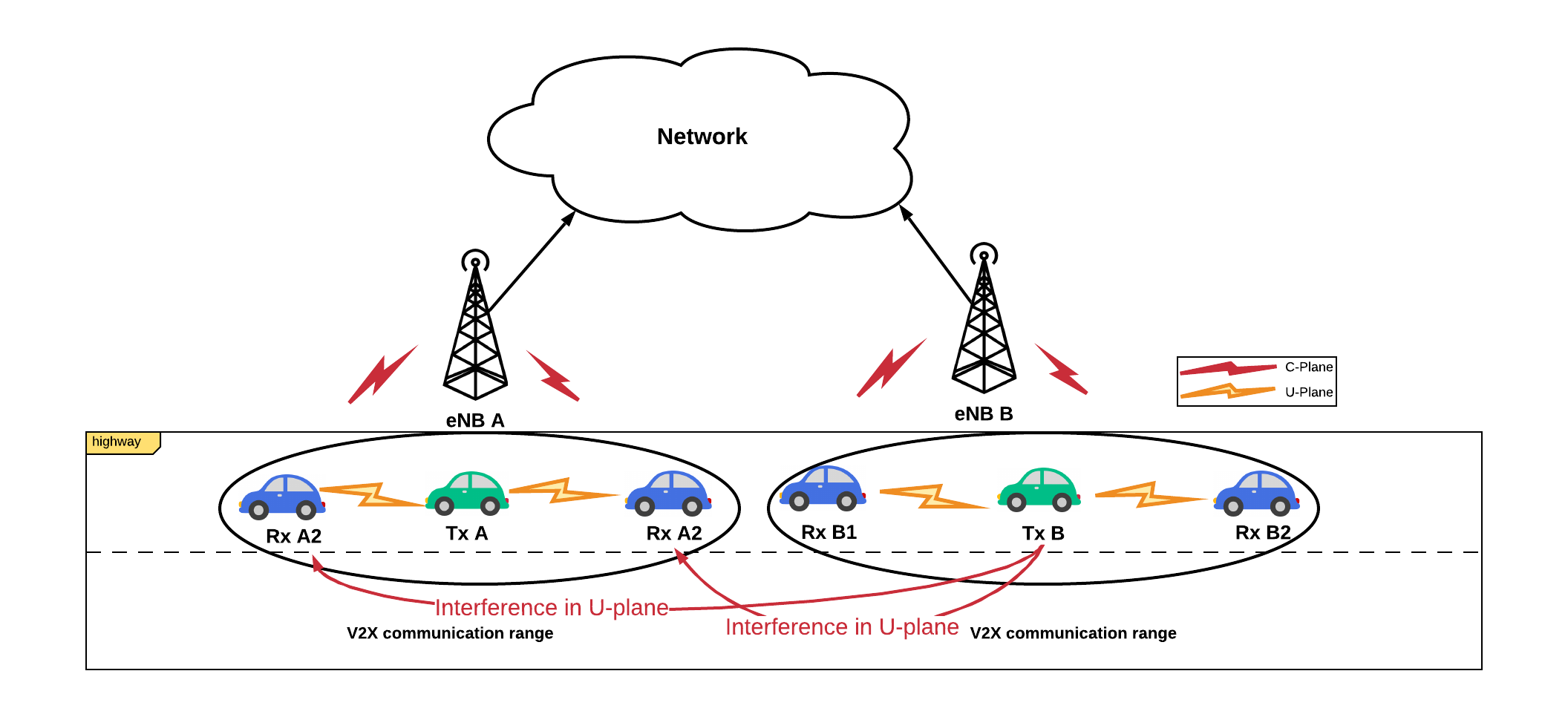}
		\caption{Direct C-V2X communication with network assistance on a highway with interference}
		\label{fig}
	\end{figure*}

The Direct C-V2X (DC-V2X)communication through sidelink is a mode of communication wherein a User Entity (UE) can directly communicate with other UEs in its proximity over the PC5 air interface proposed in 3GPP [5]. This communication is a point-to-multipoint communication where several Receivers (Rxs) try to receive the same data packets transmitted from a single Transmitter (Tx). As shown in Fig.1, a network-assisted DC-V2X transmission model is implemented in this work, using a highway scenario for data packets transmission, where all UEs are connected to Base Stations (BSs). What's moreover, the UE radio architecture consisting of the U-Plane and C-Plane is provided for C-V2X communication [6]. In this work, the Tx directly transmits its data packets to the surrounding Rxs in the communication range of the Tx in the U-Plane. This method of DC-V2X communication results in lower latencies, as it does not involve the C-Plane. And all UEs are connected to the operator network in the C-Plane where can provide network control for the DC-V2X communication [7]. In 3GPP, there are two sidelink transmission modes to assign radio resources to C-V2X Txs. They are as follows:
	\begin{itemize}
		\item Sidelink transmission mode 3: This transmission mode is only available when the vehicles are under cellular coverage. In order to assist the resource allocation procedure at the BS, UE context information (e.g., traffic pattern geometrical information) can be reported to BSs.
	\end{itemize}

	\begin{itemize}
		\item Sidelink transmission mode 4:  In contrast to mode 3, transmission mode 4 can operate without cellular coverage. In this mode, a Tx in C-V2X communication can autonomously select a radio resource from a resource pool, which is either configured by network or pre-configured in the user device for its DC-V2X communication over PC5 interface. 
	\end{itemize}
In this work, only transmission mode 3 is utilized for the DC-V2X communication through sidelink, which means all UEs are under the coverage of the cellular network and utilize resources allocated by the BSs. In this work, Two simulators are implemented in order to inspect the performances of the communication system as introduced.

\section{Sidelink Link-level Simulatior}
As mentioned before, in order to investigate the performance of large scale wireless communication in the future, we first implement the sidelink LL simulator to evaluate the physical layer algorithm performance using a single UE. In other words, LL simulation is basically a software implementation of one or multiple links between the Tx and the Rx, with a generated channel model to reflect the actual transmission of the waveforms from Tx to Rx [4]. In this simulator, we choose Extended Vehicular A Model (EVA) according to the simulation scenario, i.e.the highway. Moreover, link-state tracking, prediction algorithm, synchronization algorithm, Multiple Input Multiple Output (MIMO) gains, and Adaptive Modulation and Coding (AMC), Rxs’ structures, modeling of channel encoding and decoding are all investigated and implemented in the LL simulatior. The required results of the Block Error Rate (BLER) and Signal to Interference plus Noise Ratio (SINR) are generated as a baseline for measuring the performance of the communication system from the LL, which is the essential part for implementing sidelink SL simulator [4].

\section{Sidelink System-level Simulatior}
 Usually, the SL simulation is used to measure the performance of cell networking and simulating large networks comprising of multiple eNBs and UEs [3]. The system relevant functionalities such as network layout, channel model, L2S mapping, and characteristics of a stationary BS and mobile UE are taken into account to provide a good insight into the SL simulator. SL simulation is generally based on multi-cell multi-user, as a result its channel considers large-scale fading at the same time. 
 For SL simulation of the sidelink C-V2X communication, data packets are directly transmitted from a Tx to the Rx without going through the network infrastructure. BSs are only responsible for providing the control information to UEs under coverage. As shown in Fig.1, the interferences are from the Txs who are utilizing the same transmission bandwidth resource and transmitting data packets simultaneously. Therefore, in SL simulation, UE capacity, scheduling, mobility management, admission control, and interference management are investigated and implemented. In the simulation-based approach, the Key Performance Indicators (KPIs) need to be derived by running a simulator taking account of the different features in the relevant protocol layers, e.g., Modulation and Coding Scheme (MCS), and Hybrid Automatic Repeat Request (HARQ). In addition, Bit Error Rate (BER), BLER, SNR, SINR, and Packet Reception Ratio (PRR) describing the system performance are all inspected and simulated in the SL simulator [8]. Moreover, a mapping table from SINR to BLER has been generated from a LL simulator and integrated into an SL simulator. Furthermore, mutual information based L2S is one of the commonly used methods which has been considered to be preferable and applicable.

\section{Procedures of LL simulator}
As shown in Fig.2, a sidelink LL processing chain has been introduced. The process chain is explained as follows:
   	\begin{itemize}
   	\item Simulation configuration
   	\end{itemize}
 \begin{table}[htbp]
 	\caption{resource pool framework}
 	\begin{center}
 		\begin{tabular}{|c|c|}
 			\hline
 			LL Parameters                       & values   							 \\ 
 			\hline
 			Bandwidth resource                  & 10 MHz    							 \\ 
 			\hline
 		    Subcarrier         		            & 15 KHz      						      \\ 
 			\hline 
 			Size of PRB                         & 12 subcarriers                          \\
 			                                    & 180 KHz							      \\
 			\hline
 			Number of PRBs          		    & 48       								  \\ 
 			\hline
 			Size of SCI                         & 2 RBs                                   \\
 			\hline
 			Size of subchannel                  & vary number of PRBs      				  \\ 
 			\hline
 			Velocity                            & 100 or 260 or 350      				  \\ 
 			                                    & 400 or 450 or 500   km/h                \\
 			\hline
 			Subframe                            & 1 ms  						          \\ 		
 			\hline
 		\end{tabular}
 	\end{center}
 \end{table}  
In Tab.I, the simulation configuration is set which consists of parameters such as allocated 10 MHz channel bandwidth, allocated 48 Physical Resource Blocks (PRBs) for the sidelink UE transmission along with the 180 KHz transport block size and the Quadrature Phase Shift Key (QPSK) method. In the channel configuration, we take Doppler effect into consideration due to the highway scenario[9]. 
   	\begin{itemize}
	\item Resource pool selection
    \end{itemize}
The resource pools are created as per the allocated resource blocks and are mapped for transmission. The PRBs are selected from the available PRBs. PRBs are organized in groups called "subchannels". Moreover, the frequency-based resource allocation is configured with subchannel granularity. 
   	\begin{itemize}
	\item Control channel processing 
	\end{itemize}
For C-V2X transmission, a new Sidelink Control Information (SCI) message is generated that consists of information such as the MCS, resource indication value, time resource pattern, group destination identity, frequency hopping flag, and re-transmission opportunity for informing receiving C-V2X UEs, noted that in format 1 [6]. The generated binary message is first attached with a 16-bit Cyclic Redundancy Check (CRC) and then encoded by using a convolutional encoder, following up by interleaving and then rate matching. After generating the binary codeword, the next processing steps involve Physical Sidelink Control Channel (PSCCH)-specific scrambling, QPSK modulation, and Single Carrier-Frequency Division Modulation Access (SC-FDMA) transform precoding to generate symbols. Finally, the generated  Physical Sidelink Shared Channel (PSSCH) symbols are cyclic shifted with a random value chosen from {0,3,6,9}. Demodulation Reference Signals (DMRSs) are useful for coherent demodulation where 4 DMRS symbols are transmitted in each subframe at the indices {2,5,8,11} on two consecutive resource blocks [6]. 
	\begin{figure}[htbp]
	\centering
	\includegraphics[width=\linewidth]{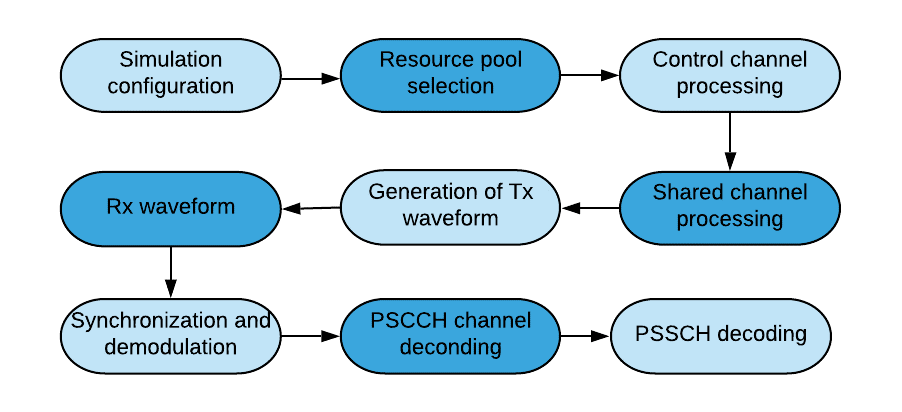}
	\caption{C-V2X LL transmitting and receiving Chain}
	\label{fig}
	\end{figure} 
   	\begin{itemize}
	\item Shared channel processing
	\end{itemize}
The PSSCH transport channel processing includes type-24A CRC calculation, code block segmentation (including type-24B CRC attachment, if present), turbo encoding, rate matching with redundancy version, code block concatenation, and interleaving.  The generated codeword is scrambled, modulated using QPSK and transform precoded in order to generate data symbols [6]. Similar to the control channel, DMRS symbols are added and transmitted alongside the data symbols [6] using the two Single Carrier-Frequency Division Multiple Access (SC-FDMA) symbols allocated to DMRS in a PSSCH subframe. PSCCH and PSSCH for C-V2X transmission may be assigned adjacent PRBs. Specifically, each PSCCH is loaded in two consecutive PRBs in a single subframe. PSSCH PRB allocation may start at the next available PRB of the pool after PSCCH is considered, or at any other PRB for the non-adjacent mode.
   	\begin{itemize}
	\item Generation of Tx waveform 
	\end{itemize}
All the symbols are then mapped to the sidelink resource grid followed by SC-FDMA modulation to create the time domain waveform. 
   	\begin{itemize}
	\item Generation of Rx waveform 
	\end{itemize}
The generated time domain waveform is then filtered through a channel. Then, the Rx waveform has been received. 
   	\begin{itemize}
	\item Synchronization and demodulation
    \end{itemize}
For each resource pool, as configured in the resource pool selection, the receiver tries to perform a blind decoding of the control information by iterating over all possible cyclic shift values. For each selected cyclic shift, the receiver first corrects the frequency offset, demodulates the SC-FDMA time domain symbols to recover the resource grid. This is followed by channel estimation method using cubic interpolation over a pre-specified time and frequency windows. The effect of the channel is equalized by dividing the received grid with that of the estimated channel grid. After this, the control symbols are extracted and then decoded by performing the inverse operations to recover the SCI message.
   	\begin{itemize}
	\item PSCCH decoding
	\end{itemize}			
After this, the control symbols are extracted and are then decoded by performing the inverse operations to recover the SCI message which contains control information. If the control information decoding is successful, the data message is to be decoded as next, whilest the unsucessful decoded data is discarded.
	\begin{itemize}
	\item PSSCH decoding
	\end{itemize}
After decoding the SCI message and the recovering, the CRC mask value is needed for performing the CRC check of the decoded data block. After recovering the data symbols, similar operations like channel estimation, equalization, and turbo decoding are performed to recover the data blocks. Finally, the BLERs are calculated by checking the CRC flags of both the control and data blocks [7].
	\begin{itemize}
		\item Channel model
	\end{itemize}
The propagation effects largely influence the quality of the transmission. Therefore, EVA has been applied in this work according to the scenario [10].
	\begin{itemize}
		\item KPI
	\end{itemize}
For sidelink simulation, we generate the mapping table from BLER to SINR as its KPI.
\section{Procedures of SL simulator}
In order to evaluate the sidelink C-V2X communication, the SL simulator has been implemented to inspect on the performance of the communication. We provide a detailed system-level simulation chain and parameters in Fig.3. Detailed information and guidelines of other systems is provided in [6].
	\begin{figure}[htbp]
		\centering
		\includegraphics[width=\linewidth]{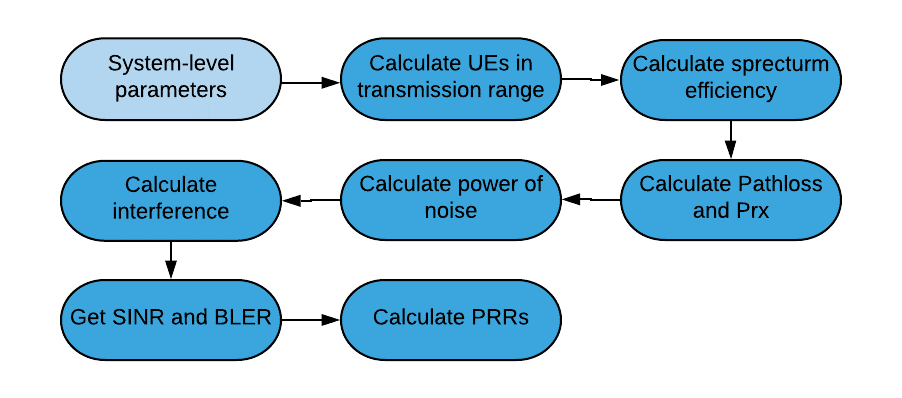}
		\caption{C-V2X SL simulator procedures}
		\label{fig}
	\end{figure} 
\subsection{Simulation Assumptions}
There are plenty of parameters utilized in the SL simulator.
	\begin{itemize}
		\item Environment model
	\end{itemize}
In Tab.II, BSs are deployed with an Inter-Site-Distance (ISD) of 1732 meters alongside the 3-kilometer highway scenario with 3 lanes in each direction to provide the network connections to the UEs of C-V2X communication. We assume the communication range of the UEs is 400 meters [8].
	\begin{table}[htbp]
	\caption{General simulation parameters of system-level simulator}
	\begin{center}
		\begin{tabular}{|l|l|}
			\hline
			SL parameters & Values\\ 
			\hline
			Scenario & Highway \\ 
			\hline
			Cellular layout & ISD 1732m  \\ 
			\hline
			Length of Highway  & 	3000 meters  \\
			\hline
			Number of lanes  & 	6   \\
			\hline
			Lane width & 	4 meters    \\
			\hline
			Antenna height of BS  & 	35 meters   \\
			\hline
			Antenna height of UE  & 	1.5 meters   \\
			\hline
			Channel bandwidth  & 	10 MHz   \\
			\hline
			Carrier frequency & 	5.9 GHz   \\
			\hline
			Data traffic  & 	Highway: 10 Hz (CAM)   \\
			\hline
			Packet size  & 	256 bytes   \\
			\hline
			Tx power & 	24 dBm   \\
			\hline
			Required communication range  & 	400 meters   \\
			\hline
			Required PRR  & 	$\geq$ 99 \%     \\
			\hline
			Vehicle density or IVD  & 	54 vehicles/cell or 100 meters   \\
			\hline
			Noise figure & 	9 dB   \\
			\hline
			Channel model & 	WINNER II channel model   \\
			\hline
			Tx antenna gain	 & 0 dB   \\
			\hline
			Rx antenna gain	 & 3 dB   \\
			\hline
		\end{tabular}
	\end{center}
\end{table}
	\begin{itemize}
		\item Deployment model
	\end{itemize}
In this work, the antenna height of BS utilized is 35 meters [8]. An isotropic antenna is installed on the top of each vehicle at a height of 1.5 meters. 1×2 antennas configuration is exploited for the DC-V2X communication. Also, each DC-V2X Tx has a constant transmission power of 24 dBm. Moreover, the central carrier frequency is 5.9 GHz with a transmission bandwidth of 10 MHz.
	\begin{itemize}
		\item Traffic model
	\end{itemize}
Here, we use a periodic package transmission of 256 Bytes with 10 Hz periodicity for each vehicle [12].
	\begin{itemize}
		\item Channel model
	\end{itemize}
The WINNER II channel models can be used in SL performance evaluation of wireless systems. Therefore, the WINNER II models are applied for evaluating the performance of the sidelink C-V2X communication. The WINNER II channel models [13] are the propagation models for calculating the pathloss. In this work, we utilized the WINNER II channel models to inspect on C-V2X communication.
\subsection{Calculations}
All parameters utilized in the DC-V2X communication have been provided as follows:
	\begin{itemize}
		\item Number of UEs
	\end{itemize}
As shown in Fig.1, we take the Tx A under the control of eNB A as an example. The number of Rxs in the communication range of Tx A is calculated.
	\begin{itemize}
		\item Modulation and coding scheme
	\end{itemize}
Spectral Efficiency (SE) is calculated as:
	\begin{equation}
	SE=P\times UE\times Period /BW \label{eq}
	\end{equation}
Where the $P$ is the packet size and $UE$ is the number of the utilized users. The $Period$ is the packet transmission frequency. The $BW$ is the system transmission bandwidth. In this work, we use QPSK as the de-/modulation scheme and a value of 1/3 coding rate in both LL simulator and SL simulator. Therefore, the number of UEs supported by the system is calculated as:
	\begin{equation}
	UE_{supported}=SE\times BW/ (P\times Period) \label{eq}
	\end{equation}
	\begin{itemize}
	\item Received power of a Rx
	\end{itemize}
In this work, only large-scale fading like pathloss and shadow fading are taken into consideration. The received power of the Rx in the communication range of the Tx is calculated as: 
	\begin{equation}
	P_{Rx}=P_{Tx} + G_{Rx_a} + G_{Tx_a} - PL
	\end{equation}	
Where $P_{Tx}$ is the transmission power (dBm), $G_{Rx_a}$ and $G_{Tx_a}$ (dBi) are the antenna gains of receiver and transmitter respectively. $PL$ is the pathloss of the Rx which can be calculated from the WINNER II channel mode [13].
	\begin{itemize}
		\item Noise figure
	\end{itemize} 
Noise power in an Rx is usually dominated by the thermal noise [12]. Thermal noise Power Spectral Density (PSD) is -174 dBm/Hz. The experienced noise power of the received signal can be calculated as:
	\begin{equation}
	P_{N}=Thermal_{N} + NF + BW
	\end{equation}	
where $NF$ is the noise figure of the Rx, and $BW$ is the logarithm of the reception radio bandwidth resource.
	\begin{itemize}
		\item Interference 
	\end{itemize}	
For sidelink C-V2X communication, data packets are directly transmitted from a Tx to the Rx without going through the network infrastructure. BSs are only responsible for providing the control information to UEs under coverage. Interference is from the Txs who are utilizing the same transmission bandwidth resource and transmitting data packets simultaneously. In Fig.1, Both Tx A and Tx B are using the same frequency resource for data packets transmission. Rx A1 and Rx A2 are in the communication range of Tx A. Thus, interference coming from Tx B can be experienced by Rx A1 and Rx A2, which try to receive the data packets from Tx A.
	\begin{itemize}
		\item SINR and BLER
	\end{itemize}	
The SINR values of the Rxs are calculated as:
	\begin{equation}
	SINR= \frac {P_{R_x}} {P_{Inter}+P_{Noise}}
	\end{equation}
Where $P_{Rx}$ is the receiving power of the wanted Tx, the $P_{Inter}$ and $P_{Noise}$ are the powers of the interference and noise respectively. From the LL simulator, we get the L2S mapping table from SNR to BLER as shown in Fig.4.
	\begin{itemize}
		\item Packet reception ratio
	\end{itemize}
PRR is defined as a percentage of nodes that successfully receive a packet from the tagged node among the Rxs that are within transmission range of the Tx at the moment that the packet is sent out [13]. In this work, a BLER value of 1\% is set as the threshold. Tx searches for all Rxs which have a calculated BLER less than a pre-defined threshold value. Then, we get the PRR values to represent the system performance of the DC-V2X communication [14].
\section{Simulation results and analysis}
In this section, the simulation results obtained from the LL and SL simulators are provided. BLERs and PRRs are utilized to represent link performances and system performances of the sidelink C-V2X communication on the highway scenario respectively.
\begin{itemize}
	\item Sidelink LL simulation result
\end{itemize}
As described before, we build the sidelink LL simulator and the simulations are run with different vehicle velocities. As shown in Fig.4, the simulator produces BLER-to-SNR curves for different velocities used in LL simulator. In this work, we use QPSK and the coding rate is fixed to 1/3. The simulation results are based on the mobile velocities of the UEs. As we know, with the increased speed, the Doppler effect on the channel propagation will fiercely be raised. For example, when we apply 0 dB of SNR, the BLER value will be increased from 3.78 $\%$ to 99.99 $\%$ from 100 km/h to 500 km/h.
\begin{figure}[htbp]
	\centering
	\includegraphics[width=\linewidth]{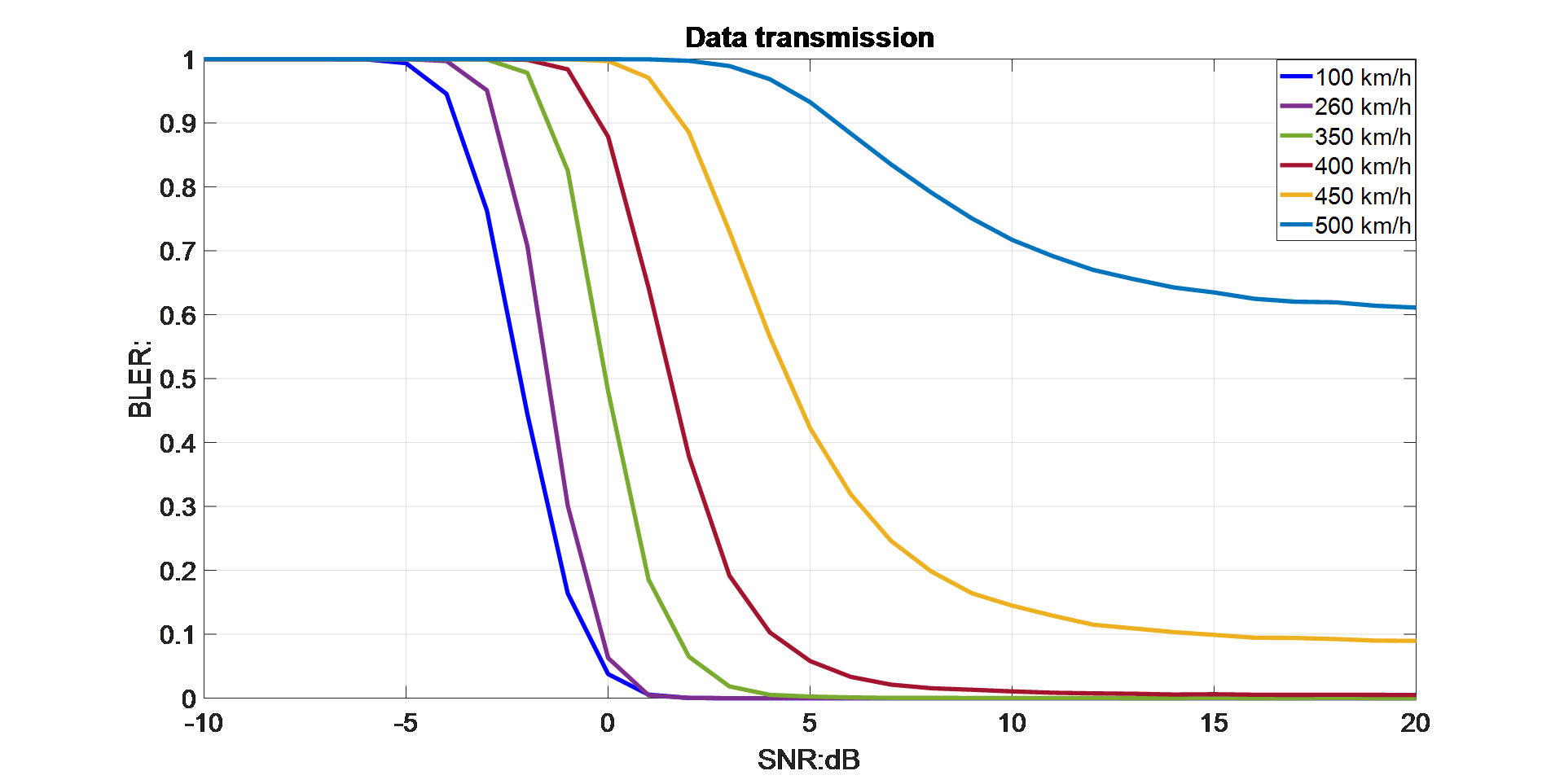}
	\caption{BLER and SNR mapping table}
	\label{fig}
\end{figure}
\begin{itemize}
	\item Sidelink SL simulation results
\end{itemize} 
\begin{table}[htbp]
	\caption{variables and functions}
	\begin{center}
		\begin{tabular}{|c|c|}
			\hline
			Variable and function    & Description        											 \\ 
			\hline
			bler$\_$i         		 & BLER value of the i-th Rx        							 \\ 
			\hline 
			counter.No        		 & get the number of Rxs which have successfully 				 \\
			                  		 & received the data packets from the Tx.        			 	 \\
			\hline
			No.UE             		 & the number of Rxs in the communication range                  \\
			                         & of the Tx.                                                    \\
			\hline
			sinr$\_$i          		 & SINR value of the i-th Rx       								 \\ 
				\hline
			UE$\_$supported          & the number of Rxs can be supported by the system    		     \\ 
			\hline
	     	velocity                 & velocity of UEs in the communication range      				 \\ 
			\hline
	        BLER$\_$SINR$\_$MCS      & get BLER value of the i-th Rx with      						  \\ 
            (SINR,velocity)          & given SINR value of the i-th Rx and velocity                   \\
			\hline
			rand(1,1)                & randomly generate a constant between 0 and 1       			  \\ 	
			\hline
		\end{tabular}
	\end{center}
\end{table}
In sidelink SL simulator, we use PRR values to represent the system performance of C-V2X communication. Fig.5 illustrates the algorithm how to calculate the PRRs. As shown in Tab.III, variables and functions utilized in the algorithm are explained. 

Step1: The SINR value of the i-th Rx among the communication range of the Tx with fixed velocity has been calculated as mentioned in Eq.5. 

Step2: Depending on the mapping table in Fig.4 from LL simulation, the BLER value of i-th Rx is generated.

Step3: A value of 1$\%$ BLER threshold is set. If the BLER value of i-th Rx is less than the threshold, then jumping to step4. Otherwise, jumping to step2.

Step4: In this step, the random tests of BLER values are used in the SL simulator. In each loop, a random constant will be generated and it is used to compare with i-th BLER. If the BLER value of i-th Rx is less than the random constant, then jumping to step5. Otherwise, return to step2.

Step5: Keeping summing the number of Rxs which have successfully received the data packets from the Tx.

Step6: The ratio of the total number of Rxs which have successfully received the data packets from the Tx  generated from step5 and the number of Rxs in the communication range of Tx is calculated. Then, an another ratio has been calculated between the number of UEs supported and the number of Rxs in the communication range of Tx. The product of the two ratios is the PRR value which are used to represent the system performance of C-V2X communication. 
\begin{figure}[htbp]
	\centering
	\includegraphics[width=\linewidth]{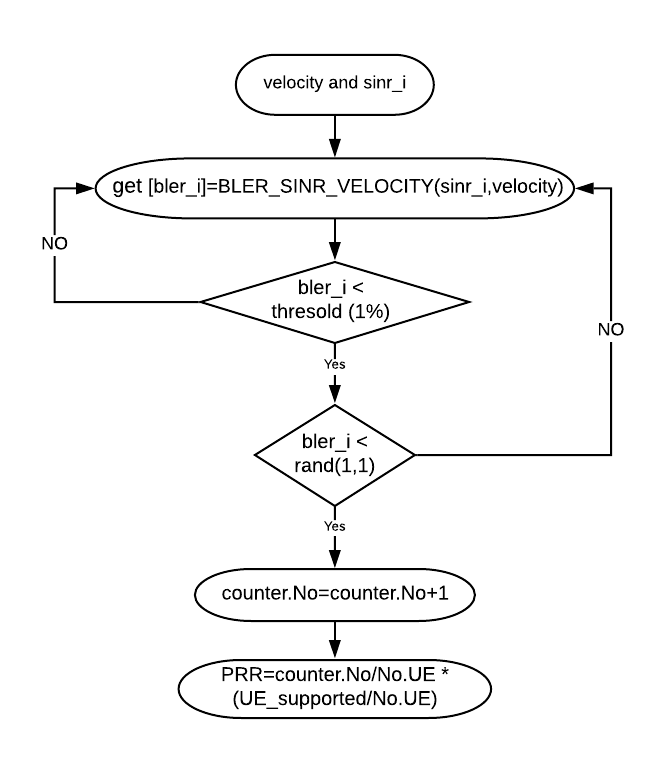}
	\caption{Algorithm of calculating PRR values}
	\label{fig}
\end{figure}
\begin{figure}[htbp]
	\centering
	\includegraphics[width=\linewidth]{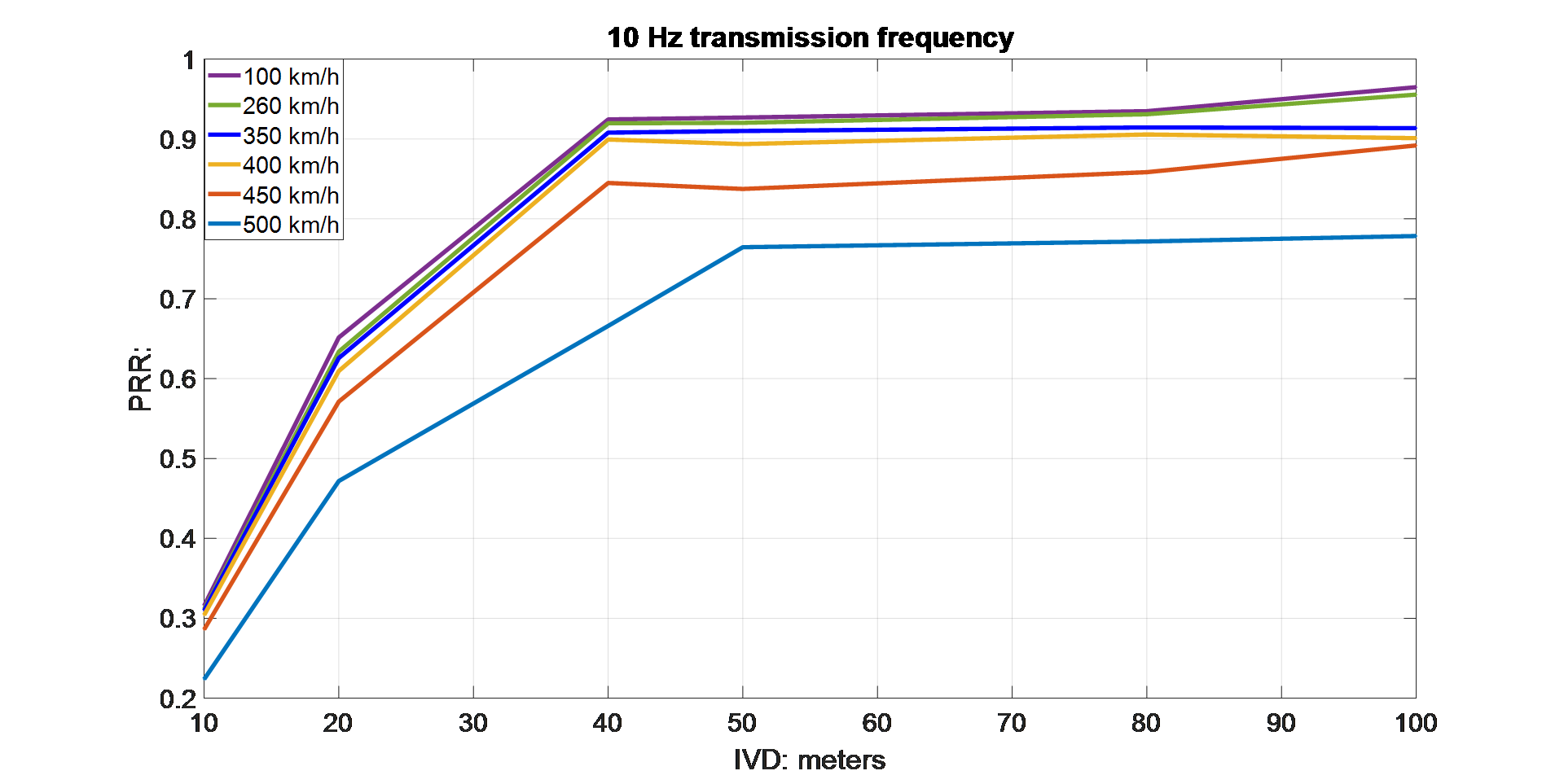}
	\caption{PRRs of different IVDs and mobile velocities with 10 Hz transmission frequency}
	\label{fig}
\end{figure}  
\begin{figure}[htbp]
	\centering
	\includegraphics[width=\linewidth]{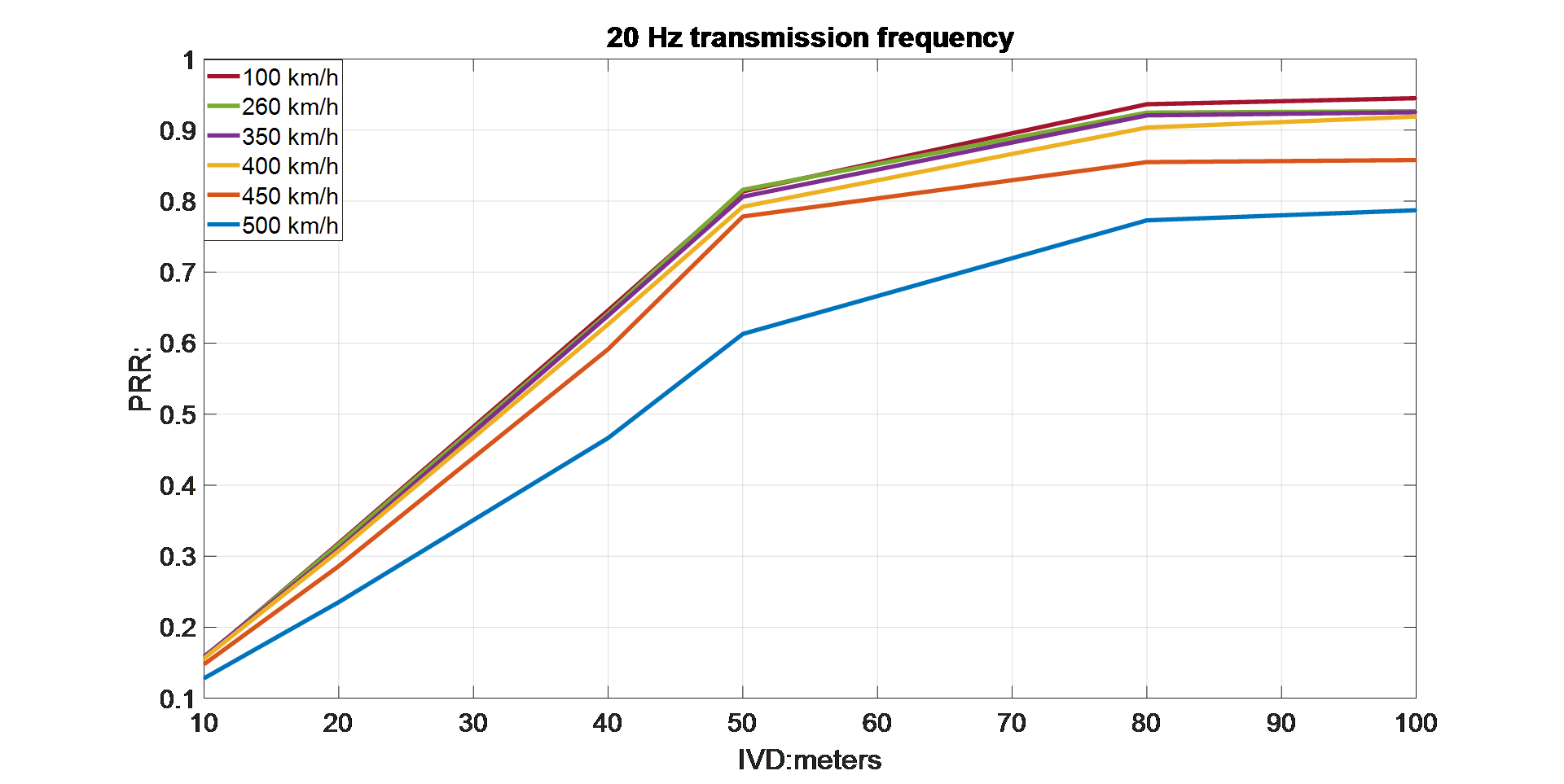}
	\caption{PRRs of different IVDs and mobile velocities with 20 Hz transmission frequency}
	\label{fig}
\end{figure}  
\begin{figure}[htbp]
	\centering
	\includegraphics[width=\linewidth]{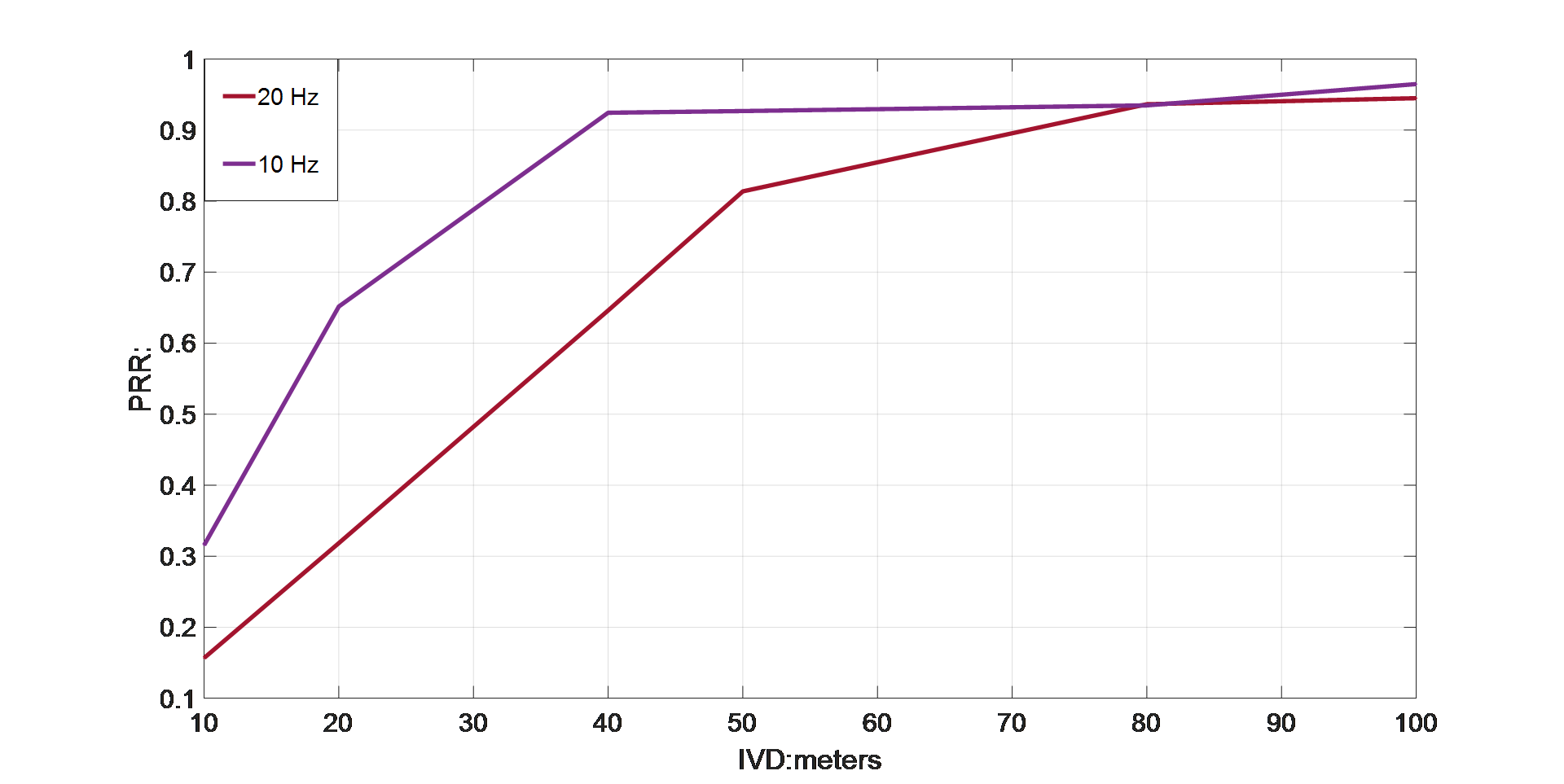}
	\caption{PRRs of different IVDs and transmission frequency with 100 km/h}
	\label{fig}
\end{figure}  

Based on the mapping table of BLER and SNR we obtained in the LL simulator, we are continuing to generate the SL simulator.  System performances of different C-V2X communication schemes are provided w.r.t. PRRs. In Fig.5, all six curves represent different PRRs of LTE sidelink C-V2X communication of different Inter Vehicle Distancess (IVDs) at 10 Hz transmission frequency, which also demonstrates the influence of the different velocities of UEs. It can be clearly seen that the PRR value increases with increasing IVD. For example, when all UEs are moving with a velocity of 100 km/h, the PRR increases from 31.56$\%$ to 96.49$\%$ with the IVD increasing from 10 meters to 100 meters. An increased IVD value means fewer vehicles will be deployed on the highway scenario since a lower number of vehicles introduces a lower system load. 

As shown in Fig.6, we can also see that the PRR has decreased with increased mobile velocities because of the Doppler effect. With a higher velocity, the Doppler effect on communication will also become larger. For instance, for a sidelink DC-V2X communication with a IVD value of 100 meters, we get the PRR value of 94.50$\%$  when the velocity is 100 km/h. However, if we increase the velocity to 500 km/h, then the PRR value is 78.72 $\%$. 

As shown in Fig.7, we get two curves at different transmission frequencies for the DC-V2X communication with 100 km/h speed. It's easy to note the PRR values increase when vehicles are transmitting with a decreased transmission frequency. For example, when we apply an IVD of 10 meters, the PRR is increasing from 15.67$\%$ to 31.56$\%$ at 10 Hz transmission frequency compared to 20 Hz. The reason for this increase is that, in Eq.2, if we reduce the transmission frequency, the number of UEs can be supported will be increased too.
\section{Conclusion}
The implementations of sidelink LL and SL simulators for the DC-V2X communication on the highway scenario are introduced in this work. We are trying to make both methodologies of LL and SL simulators easier to follow for other researchers. Different simulation objectives can be met by using different LL or SL simulation types. In LL simulator, we get the mapping table from SNR to BLER which shows the LL performance. We have applied the mapping table to the SL simulator. In the SL simulator of the DC-V2X communication on this highway, we implement the whole network to inspect on the SL performance of the DC-V2X communication. And we have provided a detailed analysis of the effect of applying different the transmission frequencies and IVDs for C-V2X communication system. In addition, we utilize different velocities for analyzing the Doppler effect on C-V2X communication. After evaluating the simulation results, we find that the system performance can be improved by decreasing the UEs or by reducing the velocities of vehicles. Last but not least, our next work is to implement more LL and SL simulators for more complicated scenarios.   

\end{document}